\documentclass[useAMS,usenatbib]{mnras}
\usepackage{graphicx}
\usepackage{amsbsy,amsmath}
\usepackage{graphics,graphicx}
\usepackage{subfloat}
\usepackage[compatibility=false]{caption}
\usepackage[skip=0.1pt]{subcaption}
\usepackage[T1]{fontenc}
\usepackage{ae,aecompl}
\usepackage[compact]{titlesec}
\def\RRI{$^{1}$}
\def\NCRA{$^{2}$}

\title[Relics in A168]{\bf Twin radio relics in a near-by low-mass galaxy cluster Abell 168}
\author[Dwarakanath et. al.]{{
K.~S.~Dwarakanath\RRI\thanks{dwaraka@rri.res.in},
V.~Parekh\RRI, R.~Kale\NCRA, L. ~T. ~George\RRI}
\\
$^{1}$Raman Research Institute, C V Raman Av, Sadashivanagar, Bangalore 560080, India\\
$^{2}$National Center for Radio Astrophysics,Tata Institute of Fundamental Research, Post Bag 3, Pune 411007, India \\}
\begin{document}
\label{firstpage}
%\author{}
\date{}
\maketitle

\begin{abstract}

\par 
We report the discovery of twin radio relics in the outskirts of the low-mass 
merging galaxy cluster Abell 168 (redshift=0.045).  One of the relics is elongated with a
linear extent $\sim$ 800 kpc, a projected width of
 $\sim$ 80 kpc and is located $\sim$ 900 kpc
toward the north of the cluster center, oriented roughly perpendicular to the 
major axis of the X-ray emission.
The second relic is ring-shaped with a size $\sim$ 220 kpc
and is located near the inner edge of the elongated relic at a distance of $\sim$ 600 kpc
from the cluster center. These radio sources were imaged
at 323 and 608 MHz with the Giant Meterwave Radio Telescope and at 1520 MHz with the
Karl G Jansky Very Large Array (VLA). The elongated relic was detected at all the frequencies
with a radio power at 1.4 GHz of 1.38$\pm 0.14 \times 10^{23}$ W Hz$^{-1}$ having a power law
in the frequency range 70 - 1500 MHz (S$\propto \nu^{\alpha}, \alpha = -1.1 \pm 0.04$).
This radio power is in
good agreement with that expected from the known empirical relation between the 
radio powers of relics and
the host cluster masses. This is the lowest 
mass (M$_{500}$ = 1.24$\times$10$^{14}$ M$_{o}$) cluster in which relics due to 
merger shocks are detected.
The ring-shaped relic has a steeper spectral index ($\alpha$) of -1.74$\pm$0.29 in the frequency
range 100 - 600 MHz.  
We propose this relic to be an old plasma revived due to 
adiabatic compression by the outgoing shock which produced the elongated relic. 
\end {abstract}

\section{Introduction}
 
Diffuse radio emission in galaxy clusters can be broadly classified into two
categories --  halos and relics, found predominantly in the central regions and
towards the peripheries of clusters respectively. Neither of them are 
associated with currently active galaxies of the clusters. Halos are typically
of $\sim$ 1 Mpc extent, have a roundish morphology, of low surface brightness
($\sim$ 1 mJy arcmin$^{-2}$ at 1.4 GHz) and are unpolarized to a few percent
level. Halos tend to have steep radio spectra, with $\alpha < -1$ , where
S $\propto \nu^{\alpha}$, S being the flux density. Relics are similar
to halos in terms of their extents in one direction with narrow ($\sim$ 0.1 Mpc)
extents in the perpendicular direction. Relics have somewhat higher
surface brightnesses, but have similar spectra as the halos. However,
they are located toward the cluster periphery ($\sim$ 1 Mpc from the cluster center)
and are polarized to the extent of $\sim$ 20 \% \citep{2012A&ARv..20...54F, 2014IJMPD..2330007B}. 
About 30 \% of
the high X-ray luminosity (L$_{x} >$ 5$\times 10^{44}$ erg s$^{-1}$ ) clusters
host radio halos and relics. A positive correlation between the radio powers
of halos and relics at 1.4 GHz and the X-ray luminosities of respective host
clusters has been observed \citep{2012A&ARv..20...54F}. 

The radio emission from halos and relics is known to be synchrotron in 
origin with $\sim$ Gev electrons radiating in the $\sim \mu$G magnetic
fields in clusters.   
The most important puzzle with the existence of such diffuse radio 
emission in galaxy clusters has been their extents. The diffusion time
scales of the relativistic electrons to cover these extents
are more than 10 times larger than the life times of the radiating electrons
at a frequency $\sim$ 1 GHz \citep{1977ApJ...212....1J}. 
One of the ways to get around this problem
is to postulate in-situ acceleration of relativistic electrons, thus 
avoiding the time scale problem. Magnetic fields have been found to be
ubiquitous in galaxy clusters \citep{2002ARA&A..40..319C}.

Cluster-wide turbulence is believed to be responsible for in-situ
acceleration of relativisitc electrons that produce the radio halos.
Cluster mergers have been invoked to create turbulence in the cluster
volume. Mildly relativistic seed electrons present in the cluster volume
(from previous AGN activities, for e.g.) 
are believed to be accelerated to relativistic energies through their
interactions with the turbulence. Turbulent acceleration is inefficient
and short-lived, as a consequence of which the halo emission is believed
to be correlated to recent merger activities of clusters. Indeed, a
correlation between the dynamical states of clusters and the existence
of radio halos and relics in them has been observed with the halos and
relics being predominantly found in non-virialised clusters \citep{2007ApJ...670L...5B, 2013MNRAS.429.3564D, 2009A&A...507..661B, 2010ApJ...721L..82C}.

At least two types of radio relics are known to exist in clusters.
The large ($\sim$ Mpc) relics are found toward the peripheries of clusters,
oriented perpendicular to the merger axes of clusters \citep{2006Sci...314..791B,
2010Sci...330..347V, 2012A&A...546A.124V}. These relics are 
produced by the out-going merger shocks. While the overall machanism for
producing the relativistic electrons is believed to be diffusive shock
acceleration at the sites of these shocks, the seed electrons can come from
both the thermal pool of electrons and the mildly relativistic fossil electrons
left over from earlier AGN activities \citep{1998A&A...332..395E, 2005ApJ...627..733M}.
Cluster merger simulations have been carried out to
account for the occurrence of relics and their properties. Some of
the basic properties of relics like their morphology, spectral index
variations from the shock edge towards the post-shock regions, and
the steepening of spectral index at higher frequencies have been
accounted for to some extent \citep{2002MNRAS.331.1011E, 2015IAUGA..2227971K, 2017JKAS...50...93K}.  A recent example of re-acceleration of 
electrons by shocks in galaxy clusters has been found in Abell 3411-3412 
\citep{2017NatAs...1E...5V}.
The second type of relics are the radio 'phoenices' that are believed to trace
the compressed fossil plasma. These 'phonecies' are found both in the central
and the peripheral regions of clusters and display steep and curved spectra
\citep{2001AJ....122.1172S, 2001A&A...366...26E, 
2002MNRAS.331.1011E, 2012ApJ...744...46K}.

\begin{figure}
\includegraphics[width=1\columnwidth]{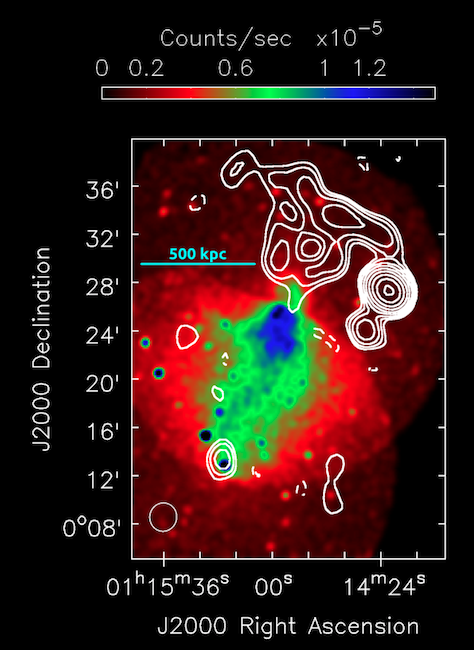}
\caption{Radio image at 200 MHz (contours) from the wideband (170-231 MHz) GLEAM survey made using the MWA 
 overlaid on the XMM-Newton X-ray image. The radio image has a synthesized beam
 of 141$^{''}$ x 134$^{''}$, -10$^{\circ}$
with an RMS ($\sigma$) of 9 mJy ~beam$^{-1}$. 
The first contour is at 3$\sigma$ and increases in 
steps $\sqrt{2}$. Dashed lines indicate negative contours.  }
\label{MWA_X-ray}
\end{figure}

\begin{table*}
\centering
\begin{small}
\caption{X-ray properties}
\begin{tabular}{ccccccccccccccc}
\hline
\hline
Cluster & \multicolumn{3}{c}{RA(J2000)} & \multicolumn{3}{c}{DEC(J2000)} & $z$ & L$_{x}$
(0.1-2.4~ keV) &M$_{SZ}$  \\
  & hh & mm & ss & $^{\circ} $ & $' $ & $ ''$ & & (10$^{44}$~erg~s$^{-1}$) &(10$^{14}$ 
M$_{\odot}$) \\
\hline
A168  &  01 & 15 &12.0 &  +00& 19& 48       &0.0450   &0.46     &1.52 \\
\hline
\end{tabular}
\label{sample}
\end{small}
\end{table*}

The diffuse radio emission  associated with the galaxy cluster Abell 168 (hereafter A168) was 
discovered in a cross referencing of the Meta-Catalogue of X-ray detected Clusters (MCXC) 
of galaxies with the GaLactic and Extragalactic All-sky MWA (GLEAM) survey at 200 MHz 
carried out by the Murchison Widefield Array.
The MCXC is a catalogue of X-ray detected galaxy clusters  
based on the publicly available ROSAT All Sky Survey and 
serendipitous discoveries \citep{2011A&A...534A.109P}. This catalog consists of a 
total of 1743 galaxy clusters with homogenised properties 
such as redshift, 0.1--2.4 keV band luminosity $L_{500}$, 
total mass $M_{500}$, and radius $R_{500}$ available for each cluster.
The GaLactic and Extragalactic All-sky MWA (GLEAM) survey was carried out using the
Murchison Widefield Array \citep{2017MNRAS.464.1146H}. The GLEAM survey covers the entire sky
south of $\delta = + 30^{\circ}$ in the frequency range 72-231 MHz. The instantaneous
frequency coverage of MWA is 30.72 MHz and hence the above mentioned frequency range
is divided into five bands providing near contiguous coverage. In addition, this survey
also produced images at 200 MHz with a bandwidth of 60 MHz resulting in a resolution
of $\sim$ 2 arcmin and a sensitivity of $\sim$ 5 mJy ~beam$^{-1}$. This diffuse radio emission
in A168 was also detected as part of the project to catalog all diffuse radio sources
in clusters using the GLEAM survey (M. Johnston-Hollitt, private communication).
An overlay of the GLEAM 200 MHz image on the XMM-Newton image of the A168 cluster
is shown in Fig. 1. 
Some basic properties of A168 are given in Table 1. 
A168 is also detected in the Planck observations (PSZ2 G135.76-62.03) and 
it's Sunyaev-Zel'dovich (SZ) mass is of $\sim$ 1.52 $\times$ 10$^{14}$ M$\odot$
(\cite{2016A&A...594A..27P}).
\par The cosmology used in this paper is  $H_{0}$ = 70 km s$^{-1}$ Mpc$^{-1}$ $\Omega_{M}$ = 0.3 and $\Omega_{\Lambda}$ = 0.7, giving a luminosity distance of 199.3 Mpc to A168 with an 
assumed redshift of 0.045.

\section{Observations}

\subsection{GMRT Observations}
The cluster A168 was observed with the Giant Meterwave Radio Telescope (GMRT)
 in the standard 325 and 610 MHz
bands during 26-27 November, 2016. A continuous track of 8 hours was carried out at each of 
the two center frequencies  with a bandwidth of 32 MHz and 512 channels. The GMRT Software Backend
was used for these observations. Primary calibrators 3C48 and 3C147 were used along with the 
secondary gain calibrator 0059+001 at both the frequencies.  

\subsection{VLA Observations}
The Karl G. Jansky Very Large Array (VLA)  
observations of A168 were carried out on 28 April, 2017 when the array was in
the D-configuration (maximum baseline $\sim$ 1 km). The total observing time was $\sim$ 2.5 hour.
The observations were carried out in the frequency range 1-2 GHz with 16 spectral windows,
each spectral window having 64 channels. The primary and secondary calibrators used were 3C138 and 
J0059+0006 respectively.

\section{Data Analysis}

\subsection{VLA Data Analysis}
For analysing the VLA data, the Very Large Array calibration pipeline,
which does basic flagging
and calibration, was used (https://science.nrao.edu/facilities/vla/data-processing/pipeline). 
The shortest baseline from which data was available was
200$\lambda$ with the longest baseline available being $\sim$ 5000 $\lambda$ ($\sim$ 1 km).
This calibrated data was further analysed using the Common Astronomy
Software Applications (CASA) package (\citep{2007ASPC..376..127M}).
 The imaging was carried out using the
multi-scale, multi-frequency synthesis method in the task CLEAN 
which takes into account the large bandwidth
used in these observations in addition to recovering the extended emission as well
as possible. Furthermore, imaging was also carried out taking into 
account the spectral curvature across the band (Taylor terms = 2 in the task CLEAN) and the wide-field 
of view (w-projection in the task CLEAN) \citep{2008ISTSP...2..647C, 2011A&A...532A..71R}. 
Suitable number of phase and amplitude self-calibrations were carried out to get the
best images as evidenced by minimum RMS and least systematics.

Images of different resolutions and sensitivities were produced by varying the 
weighting of the visibilities (robust parameter in the task CLEAN). Varying the robust
parameter \citep{1995AAS...18711202B} from -2 to +2 changes the weighting of the visibilities from 'uniform' to
'natural'. Natural weighting produces images sensitive to extended emission, but with a
loss of resolution. Uniform weighting, on the other hand, produces images with the best
resolution, but at the loss of sensitivity to extended emission.
 The final image adopted here is the one produced with robust = 1, which
is a compromise between sensitivity to extended emission and resolution.

\subsection{GMRT Data Analysis}

The GMRT observations were analysed using the Source Peeling And Modeling (SPAM) package
(\citep{2009A&A...501.1185I}.
SPAM is an Astronomical Image Processing Software (AIPS)-based Python package that provides 
semi-automated data reduction scripts for all sub-GHz Frequencies at GMRT \citep{2014arXiv1402.4889I}.
Apart from incorporating well-tested data reduction steps like calibration, wide-field imaging
and self-calibration, SPAM also incorporates direction-dependent gain calibration and imaging
and radio frequency interference excision methods \citep{2014arXiv1402.4889I}. 
Images of A168 field were produced with different weightings,
tapers and uvranges of the visibilities included in the imaging. The GMRT observations were also
analysed using AIPS and the images produced by these two methods were compared. There
was excellent agreement between the two sets of images. 

The shortest baseline included in the imaging at 323 and 608 MHz  
was 200 $\lambda$ to be consistent with the VLA observations. This restriction also
excludes some of the shortest baseline visibilities from GMRT  
which are invariably corrupted by radio frequency
interference.  The final images used here are the ones with a value of robust = 1 in the 
imaging task CLEAN for reasons
already explained.  Since the VLA observations were carried out in the
D-configuration (maximum baseline $\sim$ 1 km), the higher resolution observations at 323 and 608 MHz
from GMRT (maximum baseline $\sim$ 25 km) 
were imaged to have the same resolutions as those from the VLA observations by a suitable
tapering of the visibilities.  All the images were corrected for the respective primary beam
attenuations. The higher resolution GMRT observations were examined for possible unresolved
sources contaminating the cluster diffuse emission.

\subsection{Flux Density Scale}

The GLEAM survey is on the flux density scale of \cite{1977A&A....61...99B} and is accurate to $\sim$ 8 \%.
The GMRT flux densities are on the \cite{2012MNRAS.423L..30S} scale. These two scales 
are in agreement to within 3 \% \citep{2017MNRAS.464.1146H}. The VLA flux densities are tied
to the \cite{2017ApJS..230....7P} flux density scale which is in agreement with the \cite{2012MNRAS.423L..30S}
scale to within 5 \%. 

\begin{figure*}
   \begin{subfigure}[t]{1\columnwidth}
       \includegraphics[width=1\columnwidth]{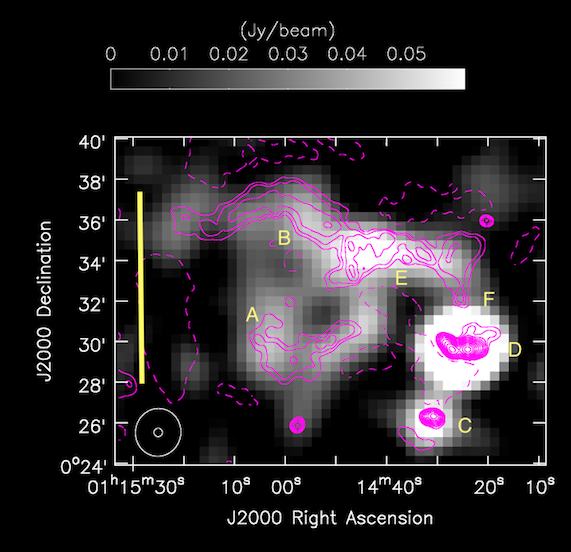}
       \caption{}
       \label{rfidtest_zaxis3}
     \end{subfigure}  
   \begin{subfigure}[t]{1\columnwidth}
       \includegraphics[width=1\columnwidth]{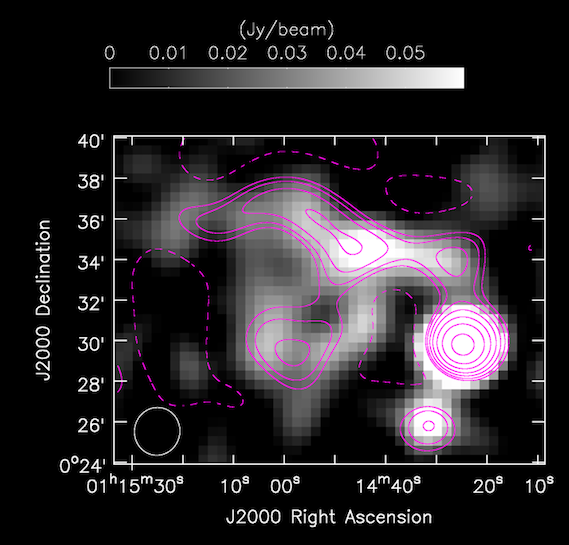}
       \caption{}
       \label{rfidtest_zaxis3}
     \end{subfigure}  
    \centering
     \begin{subfigure}[t]{1\columnwidth}
        \includegraphics[width=1\columnwidth]{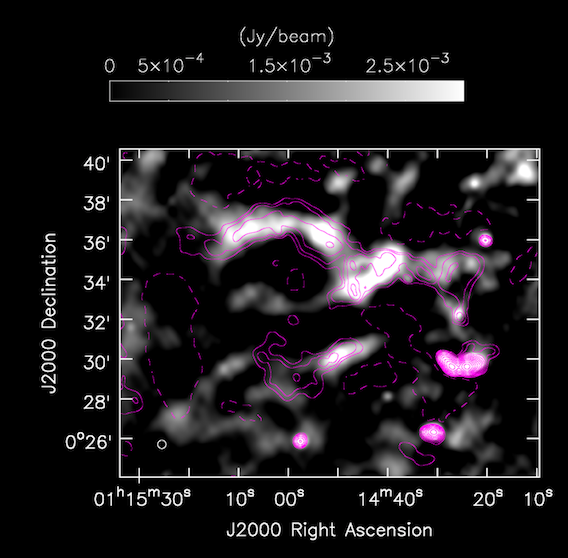}
    \caption{}
       \label{rfidtest_yaxis2}
     \end{subfigure}
   \begin{subfigure}[t]{1\columnwidth}
       \includegraphics[width=1\columnwidth]{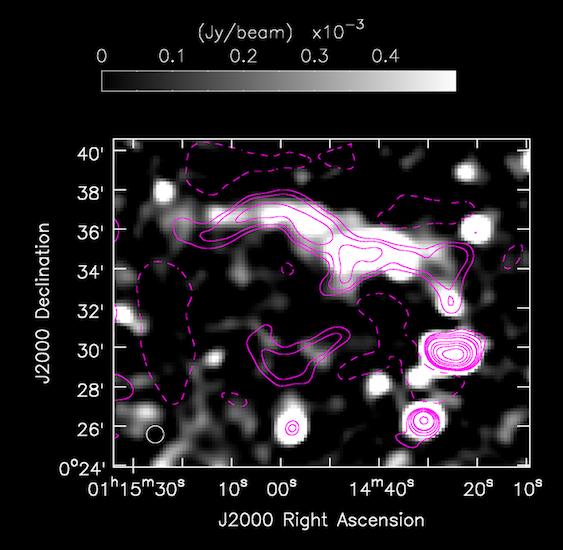}
       \caption{}
       \label{rfidtest_zaxis3}

    \end{subfigure}
      
   \caption[]{A168. (a) GMRT 323 MHz image (contours) (25$''$ $\times$ 25$''$, $\sigma$ = 1.0 mJy ~beam$^{-1}$) 
overlaid on the GLEAM 200 MHz image (greyscale). The vertical line indicates 500 kpc.
 (b) GMRT 323 MHz image convolved to the same resolution (contours, $\sigma$ = 15.4 mJy
 ~beam$^{-1}$) 
as the GLEAM 200 MHz image and
overlaid on the GLEAM 200 MHz image (greyscale) (c) GMRT 323 MHz contours 
overlaid on GMRT 608 MHz greyscale image ($\sigma$ = 0.64 mJy ~beam$^{-1}$). Both images are at a resolution of 25$''$ $\times$ 25$''$. 
(d) GMRT 323 MHz contours ($\sigma$ = 4.2 mJy ~beam$^{-1}$) overlaid on the VLA image at 1520 MHz (greyscale, $\sigma$ = 160~$\mu$Jy ~beam$^{-1}$).
 Both the images are at a resolution of 52$''$ $\times$ 52$''$. In all the images, first contour is drawn at 2$\sigma$. 
Contour levels increase in steps of $\sqrt{2}$. Dashed lines indicate negative contours. The synthesized
beams are indicated at the bottom left hand corner. }
   
   \label{A168_img}
\end{figure*}

\begin{table*}
\setlength{\tabcolsep}{4pt}
\begin{small}
\caption{Flux densities of sources A and B. The data used are from VLSSr (74 MHz, Lane et al 2014), GLEAM survey
(154 and 200 MHz), and the present work (323, 608 and 1520 MHz). The 74 MHz image at a 
resolution of 80$^{''}$ was convolved to the GLEAM survey resolution at 200 MHz. 
The integrated flux densities of sources A and B are given in mJy. 
The values in parantheses are
total errors on the flux densities estimated as described in the text.
The upper limits are 3$\sigma$ values.}
\begin{tabular}{c|cccccccccccccc}
\hline
\hline
Frequency (MHz) &   74  &      154 &   200  &   323 &  608  &  1520 \\
Synthesized beam ($^" \times ^",^\circ$)  & 141 x 134, -10 & 173 x 163, -10 & 141 x 134, -10 & 52 x 52 & 52 x 52& 52 x 52\\
RMS (mJy ~beam$^{-1}$) &200&       13.7&   8.2&      4.2&   2.0&    0.16\\
\hline
A & $<$ 600 & 339(36.6) & 149(19.3) & 96(15.8) & 28(6.6) & $<$0.48 \\
B & 864(200) &348(43) & 222(32) & 155(29.5) & 66(13.0) & 28(3.0) \\
\hline
\end{tabular}
\label{radio_prop}
\end{small}
\end{table*}

\begin{table*}
\setlength{\tabcolsep}{4pt}
\centering
\begin{small}
\caption{Discrete Sources. In case of source D, the two radio positions correspond to the
peaks of emissions of the two lobes respectively. The optical position corresponds to that
of the core between the two lobes.}
\begin{tabular}{ccccccccccccccccc}
\hline
\hline
Source & Identification & \multicolumn{6}{c}{Optical}   &$z$ & mag &  \multicolumn{6}{c}{Radio}    &Separation \\
     &  &\multicolumn{3}{c}{RA(J2000)} &\multicolumn{3}{c}{DEC(J2000)} & & &\multicolumn{3}{c}{RA(J2000)} & \multicolumn{3}{c}{DEC(J2000)} & \\
    & & hh& mm &ss  & $^{\circ} $& $ '$& $  ''$ & & & hh& mm &ss & $^{\circ} $& $ '$& $''$ &$''$ \\
\hline

C & SDSSJ011431.02+002621.4  & 01 &14& 31.0 & +00& 26& 21   & 1.76  &-     &01& 14& 31.00  & 00& 26& 19.60    & 1.8\\

D & SDSSJ011425.58+002932.7  & 01& 14 &25.6 & +00& 29 &33 &  0.35 & 20.4g &  01& 14 &27.00 & 00& 29& 39.59 & \\
  &                          &   &    &     &    &    &   &       &       & 01 & 14 &24.17 & 00& 29 &39.59& \\

E & SDSSJ011435.83+003356.9 & 01 &14 &35.8 & +00 &33& 57 &  0.54 & 21.3g  &  01& 14& 35.67 & 00 &33& 59.60  & 3.5 \\

F & CGCG385-039   &           01 &14 &25.7 & +00 &32 &10 &  0.04 & 15.5g  &  01& 14& 25.50 & 00 &32& 09.58 &  2.6 \\

\hline
\end{tabular}
\label{sample}
\end{small}
\end{table*}

\begin{figure*}
    \centering
     \begin{subfigure}[t]{0.48\textwidth}
        \includegraphics[width=1\textwidth]{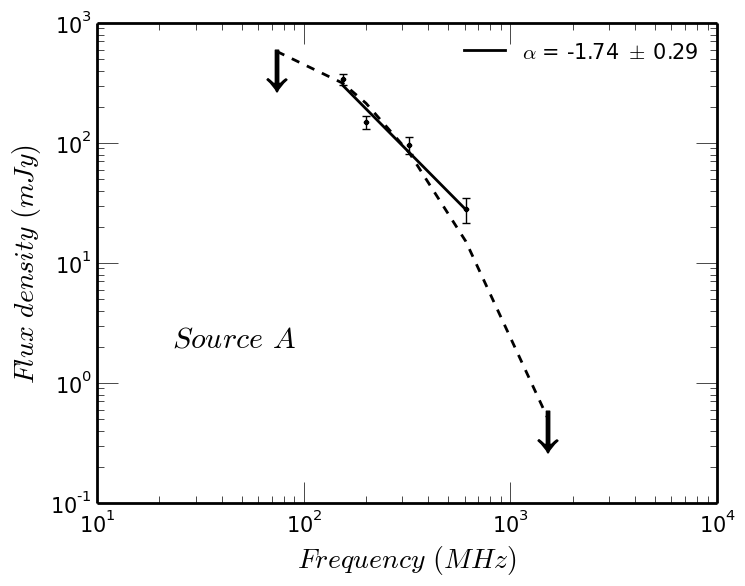}
    \caption{}
       \label{rfidtest_yaxis2}
     \end{subfigure}
   \begin{subfigure}[t]{0.48\textwidth}
       \includegraphics[width=1\textwidth]{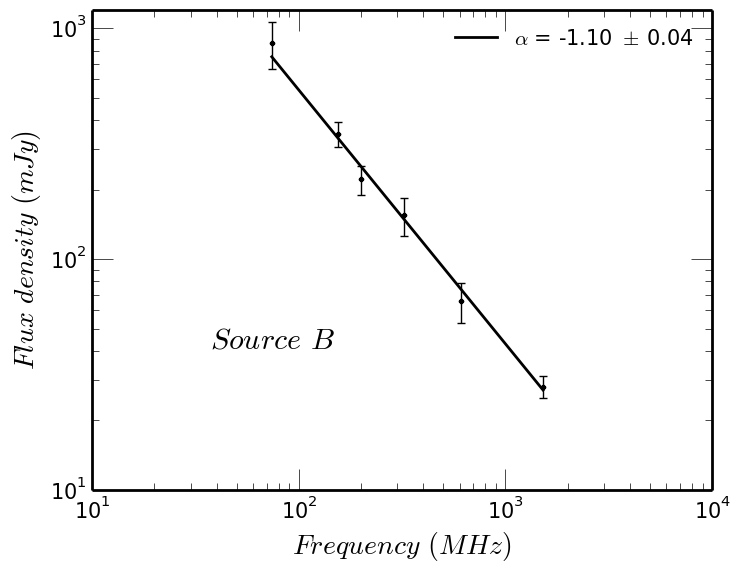}
       \caption{}
       \label{rfidtest_zaxis3}
     \end{subfigure}  
   \caption[]{Spectra of relics. A linear fit to the four data points for source A
gives a $\chi^{2}_{red}$ = 2.8 (DOF=2) with a slope 
indicated at the top right hand corner
of this panel. Given the upper limit at 1.5 GHz, the spectrum of source A is likely
to be an exponetially falling spectrum at higher frequencies as indicated by the
the broken lines. 
The best-fit power law to source B spectrum gives a 
$\chi^{2}_{red}$ = 0.4 (DOF=4) 
with a slope indicated at the top right hand corner.
The error bars correspond to total errors estimated as described in the text.}.  
\label{A168_spectra}
\end{figure*}

\begin{figure}
\includegraphics[width=1\columnwidth]{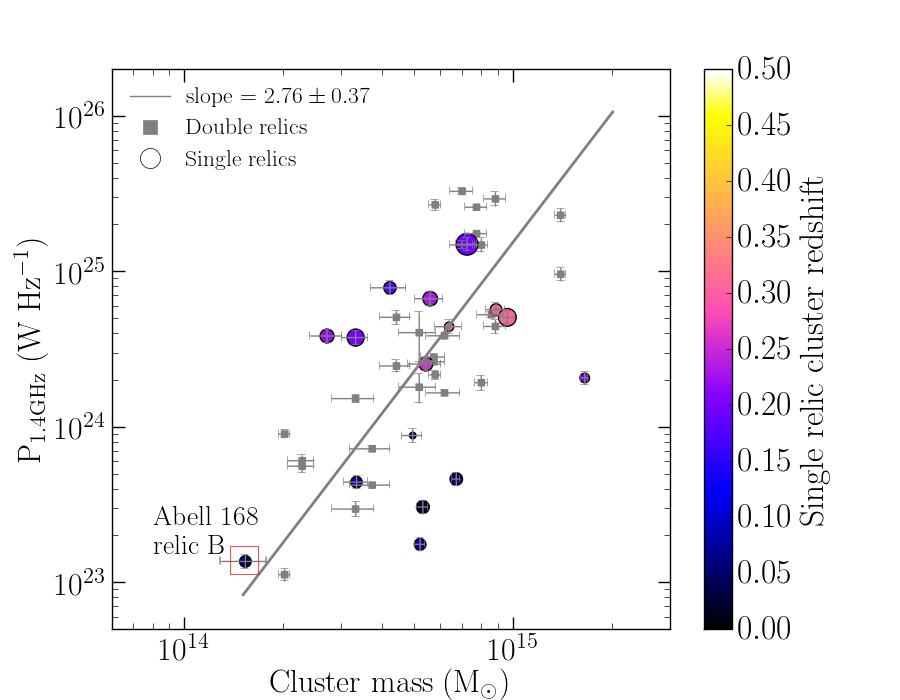}
\caption {The radio relic powers at 1.4 GHz versus the host cluster masses 
for a sample of double and single relics taken from \citep{2017MNRAS.472..940K}
 are plotted. The grey filled squares are 
the double relics for which the best-fit line is shown \citep{2014MNRAS.444.3130D}. 
The single relics are shown by circles 
filled with the colour indicating the redshift and the size scaled with 
the relic linear size.The A168 single relic (source B) is highlighted with a 
square. It is the lowest mass cluster known till date with a candidate shock 
related single relic.}
\label{power_mass}
\end{figure}

\section{Results}
An image at 323 MHz from GMRT is overlaid on the GLEAM 200 MHz image in Fig. 2a.
In Fig. 2b the above GMRT 323 MHz image convolved to the same resolution as that of the
GLEAM 200 MHz image is shown. In Fig. 2c the 323 MHz image at a resolution of 25$^{''}$
is overlaid on the GMRT 608 MHz image at the same resolution. In Fig. 2d, 
a 323 MHz image at a resolution of 52 $^{''}$ resolution is overlaid on the VLA 1520 MHz image. 

The diffuse emission detected in the GLEAM survey can be discussed as two sources
A and B, as marked in Fig. 2a. The source B refers to the northern source
while source A refers to the southern ring-shaped region. The source B is detected
in all the images shown here, while source A is detected at all except in the VLA
images. The estimated flux densities
of extended sources A and B from these images are given in Table 2.
There are two primary sources of errors in the estimation of the flux densities of
extended sources. First, there is an error due to the uncertainties in the flux 
densities of the primary and secondary calibrators at different frequencies.
This error is estimated to be $\sim$ 10\%. Second, the errors in the flux density estimates
of extended sources will be the rms in the image (Table 2) 
multiplied by the square root of 
the ratio of the solid angle of the extended source to that of the synthesized beam. 
Since these two sources of errors are unrelated, they are added in quadrature to estimate 
the total error on the flux densities of the extended sources (given in parantheses in
Table 2).

There are no flux density estimates of
sources A and B corresponding to the GLEAM survey frequencies at 87.5 and 118 MHz. 
This is because sources A and B are convolved into
one source at these frequencies due to poorer resolution. However, the sum of 
the extrapolated flux densities of sources A and B at 118 MHz is in agreement
within errors of the observed flux density of the convolved source at 118 MHz. 
The flux density estimate at 74 MHz is from
the VLSSr survey at a resolution of $\sim$ 80$^{''}$ \citep{2014MNRAS.440..327L}
convolved to the same resolution as the GLEAM survey images at 200 MHz.
The measured flux density from VLSSr was scaled down by a factor of 1.1
to be consistent with the Baars et al scale used here \citep{2014MNRAS.440..327L}.
The flux densities of sources E (S$_{323}$ = 7.7 mJy, $\alpha = - 0.89$) and 
F (S$_{323}$ = 5.4 mJy, $\alpha = -0.8$), 
which are unrelated to the diffuse source B, but which get convolved
into source B at poorer resolutions are subtracted
from the estimated flux densities of source B. In any case, 
the total flux density of sources E and F
is less than 10\% of that of source B at 323 MHz. Optical identifications for sources
C to F are given in Table 3. The spectra of sources A and B are given in Fig. 3.

\section{Discussion}
There are four discrete sources, C, D, E and F which are near-by (in projection)
to the diffuse sources A and B. As can be seen from Table 3, source D is a double radio
galaxy with an optical counterpart ( z = 0.35) in between the lobes where the core
is presumably situated. Sources C and E are background sources at redshifts of 
1.76 and 0.54 respectively. Source F is a cluster member. In the absence of any 
possible jet connecting this source to the sources A and B, source F 
does not appear related to them. Furthermore, the region of the sky covering 
the sources A and B were also searched for any optical and infra-red sources (with or 
without redshift measurement) that could
possibly have supplied fossil plasma, particularly for source A. No such sources were
detected.

The elongated (length $\sim$ 800 kpc) source B is located toward the cluster periphery
at a distance $\sim$ 700 kpc from the X-ray peak and $\sim$ 900 kpc from the cluster
center. The thickness of the relic is $\sim$ 80 kpc and the orientation of the longer
dimension of the relic is nearly perpendicular to the major axis of
X-ray emission (Fig. 1). The morphology and the location of relic B is not unlike
the well-known `sausage' and  `toothbrush' relics found in galaxy clusters
\citep{2010Sci...330..347V, 2012A&A...546A.124V}.
The spectral index
of the relic in the frequency range 70-1500 MHz is steep ($\alpha = - 1.08$, Fig. 3b), but, is well
within the known distribution of spectral indices of relics ($\alpha = -1.42 \pm$ 0.37, 
\cite{2012A&ARv..20...54F}). The relic A, on the other hand, has a  roundish, or a ring-shaped
morphology. With an extent of $\sim$ 220 kpc, it is located closer to the X-ray peak 
at a distance of $\sim$ 400 kpc. The spectrum of relic A appears curved with an 
exponential cut-off toward higher frequencies and a flattening toward lower frequencies
(Fig. 3a). The best-fit power law for the four detections between 150 and 608 MHz
implies a spectral index of $- 1.70 \pm 0.15$ (Fig. 3a). 

\subsection{Cluster dynamics and the relics}
%Galaxy cluster mergers can drive shocks and turbulence in the ICM.
The role of cluster mergers in producing shocks underlying the radio relics 
at cluster peripheries is well supported by observational evidences \citep{1999NewA....4..141G, 2006Sci...314..791B, 2009A&A...506.1083V, 2013A&A...551A..24V}. 
A different class of relics that are proposed to be fading radio galaxy lobes 
or such lobes revived due to adiabatic compression 
\citep{2001A&A...366...26E} are also well known to occur in 
clusters. Examples of relics in this category are in Abell 4038 \citep{1998MNRAS.297L..86S, 2012ApJ...744...46K} and in 
Abell 85 \citep{2001AJ....122.1172S}. In A168 we find an arc-like relic at the periphery (B) and a 
smaller steep spectrum relic (A) in its wake as seen projected in the plane of 
the sky. The ICM in the cluster is elongated along the north-south direction and 
the orientation of B is perpendicular to this direction.
Merger shock related relics have been found to preferentially be oriented 
perpendicular to the elongation axis of the ICM \citep{2011A&A...533A..35V}.
 The flat spectral index of B, arc-like morphology and the orientation thus 
 support the scenario that a cluster merger along the north-south direction 
 led to an outgoing merger shock that accelerated electrons which are detected 
 as the relic B. The relic A has a steep and curved spectrum (Fig. 3) 
 indicative of an ageing population of relativistic electrons. We propose that 
 relic A is a candidate adiabatically compressed lobe of a radio galaxy - the 
compression having been caused by the outgoing shock which is at B. 
Simulations of adiabatically compressed cocoons of radio galaxies have shown 
that as the compression proceeds, the cocoon is torn into filamentary structure 
which can appear like a single torus or multiple tori at late stages
\citep{2002MNRAS.331.1011E}. The morphology of the relic A is complex and 
filamentary which compares well with the predicted structures of compressed 
cocoons. The morphology of B is not smooth at the outer boundary but shows 
kink like features connected to A implying a possible distortion due to the 
presence of the radio cocoon in the path of the outgoing shock that led to 
the formation of A.

There is evidence for a binary merger along the north-south direction in A168 
based on optical and X-ray observations. \citet{2004ApJ...614..692Y} have 
estimated a merger timescale of 0.6 Gyr assuming the distance between two X-ray 
peaks to be 676 kpc and a colliding velocity of 600 km s$^{-1}$. 
For the relic A, if the break frequency
is considered to be at $\sim$ 100 MHz and the magnetic field is assumed to be 1 
$\mu$G, typical of the cluster outskirts
\citep{2012A&ARv..20...54F}, the life time due to synchrotron and Inverse Compton 
losses is $\sim$ 0.24 Gyr. Since the initial age of the radio cocoon which may 
have formed A is unknown, the estimated age of A could be considered a minimum. 
The time taken by a shock with a speed of 600 km s$^{-1}$ to traverse from A to 
B which is a distance of $\sim$320 kpc is 0.5 Gyr. 
These timescales are comparable to the minimum estimated merger 
timescale making the proposed scenario plausible.

Measurement of polarization in the relics A and B will provide further evidence 
for the passage of a shock. Deep X-ray observations towards the relics are also 
needed to detect the location and strength of the underlying shock.

\subsection{Implications of the low mass host cluster}
Mergers in massive systems are favoured sites for turbulent re-acceleration 
of particles leading to radio halos and shocks leading to radio relics 
\citep[e. g.][]{2015A&A...580A..97C}. Radio halos in low mass systems (M $< 
5\times10^{14}$ M$\odot$) are not common, but radio relics have been found in 
low mass systems \citep[][]{2017A&A...597A..15D, 2017MNRAS.472..940K}. We show 
the radio relic power (P$_{1.4\rm{GHz}}$) versus host cluster mass scaling with 
the relic B in A168 plotted (Fig. 4). A168 is the lowest mass system so 
far known to have radio relics of merger shock origin.  The physical mechanism 
behind the observed empirical relation between P$_{1.4\rm{GHz}}$ and 
the host cluster mass is not well understood 
\citep{2014MNRAS.444.3130D}. The sample of single relics shows a large 
scatter around the scaling, however, A168 relic B lies close to the radio power 
expected according to the scaling relation. A counter relic in A168 may be 
present but is not detected in our observations. Such counter relics have been 
found to be factors of 8 or so weaker (e. g. A3365, \cite{2011A&A...533A..35V}) and 
thus may not be detectable. The physical mechanism behind the observed 
empirical scaling is not known and the relic B in A168 extends it to even lower 
masses. This indicates that not only does the cluster mass but other factors 
such as the availability of the seed electrons, the strength of the magnetic 
field and the merger geometry itself may be playing a role in producing radio 
relics of certain radio powers. 

\section{Summary and conclusions}

We report the detection of twin radio relics in the low-mass galaxy cluster A168. 
These relics were discovered in a cross referencing of the MCXC catalogue with 
the GLEAM survey. We have imaged these relics at 230 and 610 MHz using the GMRT
and at 1520 MHz using the VLA. One of the relics (B) is elongated ($\sim$ 800 kpc),
thin ($\sim$ 80 kpc) and is located towards the northern periphery of the
cluster at a distance of $\sim$ 900 kpc from the cluster center. The second relic (A)
is ring-shaped, with an extent of $\sim$ 220 kpc and is located near the inner
edge of B. Relic B was detected at all the observed frequencies
with a radio power at 1.4 GHz of 1.38$ \pm 0.14 \times 10^{23}$ W Hz$^{-1}$ having a power law
in the frequency range 70 - 1500 MHz (S$\propto \nu^{\alpha}, \alpha = -1.08 \pm 0.04$).
This radio power is in
good agreement with that expected from the known empirical relation between the 
radio powers of relics and
the host cluster masses, with the current relic detection from the lowest 
mass (M$_{500}$ = 1.24$\times$10$^{14}$ M$_{o}$) cluster. Relic A
has a steeper spectral index ($\alpha$) of -1.70$\pm$0.15 in the frequency
range 100 - 600 MHz and appears connected to source B.
We propose source A to be an old plasma revived due to 
adiabatic compression by the outgoing shock which produced source B. The orientation
of source B, the projected distance between sources A and B, and the radiative life
time of source A are consistent with such a scenario. Polarization measurements
of source A and B and deeper X-ray observations at the positions of sources A and B 
will be needed to detect the underlying shock.

\section {Acknowledgments}

We thank Huib Intema for useful discussions on using SPAM. We thank the referee for
a critical reading of the manuscript and constructive criticisms.
We thank the staff of the GMRT that made these observations possible. GMRT is run by the National Centre for Radio Astrophysics of the Tata Institute of Fundamental Research.
The National Radio Astronomy Observatory is a facility of the National Science Foundation operated under cooperative agreement by Associated Universities, Inc.
%\newpage

\bibliography{references}
\end{document}